# Robust atmospherically stable hybrid SrVO$_3$/Graphene//SrTiO$_3$ template for fast and facile large-area transfer of complex oxides onto Si

Asraful Haque*, Suman Kumar Mandal, Antony Jeyaseelan, Sandeep Vura, Pavan Nukala*, Srinivasan Raghavan*.

Center for Nanoscience and Engineering, Indian Institute of Science, Bangalore

**Abstract:**

Heterogenous integration of complex epitaxial oxides onto Si and other target substrates is recently gaining traction. One of the popular methods involves growing a water-soluble and highly reactive sacrificial buffer layer, such as Sr$_3$Al$_2$O$_6$ (SAO) at the interface, and a functional oxide on top of this. To improve the versatility of layer transfer techniques, it is desired to utilize stable (less reactive) sacrificial layers, without compromising on the transfer rates. In this study, we utilized a combination of chemical vapor deposited (CVD) graphene as a 2D material at the interface and pulsed laser deposited (PLD) water-soluble SrVO$_3$ (SVO) as a sacrificial buffer layer. We show that the graphene layer enhances the dissolution rate of SVO over ten times without compromising its atmospheric stability. We demonstrate the versatility of our hybrid template by growing ferroelectric BaTiO$_3$ (BTO) via PLD and Pb(Zr, Ti)O$_3$ (PZT) via Chemical Solution Deposition (CSD) technique and transferring them onto the target substrates and establishing their ferroelectric properties. Our hybrid templates allow for the realization of the potential of complex oxides in a plethora of device applications for MEMS, electro-optics, and flexible electronics.

## Introduction

The family of complex oxides boasts of diverse functional materials, including piezoelectrics, ferroelectrics, materials with high dielectric constants, magnetic materials with different orders, high T$_c$ superconductors, multiferroics, and electro-optical modulators [1–6]. Epitaxial thin films of complex oxides offer a promising avenue to engineer new functions through the manipulation of strain and defects [7,8]. However, integrating these systems onto Si has been challenging, owing to the presence of amorphous native oxide [9]. One way this issue has been addressed is by using more thermodynamically stable buffer layer oxides or nitrides such as MgO, $ZrO_2$, and TiN, among others, which scavenge the native oxide inside the vacuum (deposition) chamber [10–12].

Alternatively, layer transfer techniques are viable for heterogeneously integrating complex oxides on Si and other target substrates and promise not just Si integrability but also CMOS compatibility [13–15]. This method involves transferring the functional oxide layer from the growth substrate to the target substrate. These techniques are also attractive for flexible electronics applications, which require epitaxial integration of functional oxides on flexible substrates that cannot withstand high temperatures during growth [16].

A standard layer transfer method for complex oxides employs a sacrificial layer between the top film and substrate. The sacrificial layer releases the top layer upon etching, which can then be transferred to the desired substrate. Etchants can be acidic or basic, or neutral solvents. Layers such as $La_{0.7}Sr_{0.3}MnO_3$ (LSMO), $Sr_{0.3}MnO_3$ (SMO), $SrRuO_3$ (SRO), $YBa_2Cu_3O_7$ (YBCO), MgO, ZnO [17–21] which can be etched in acidic and/or basic media suffer from the non-existence of preferential selectivity of the etchants.

By now, it has been established that a class of water-soluble oxides with lattice constants ranging from 3.8-4.1 A°, close to the lattice parameters of most functional perovskite oxides and commercially available oxide substrates, and good selectivity show immense potential as sacrificial layers [22,14]. These oxides include aluminates such as $(Ca, Sr, Ba)_3Al_2O_6$, binary oxides such as BaO, and vanadates such as $SrVO_3$ (SVO) [23–29], with aluminates being the most explored layers. $Sr_3Al_2O_6$ layers, in particular, are very reactive to moisture, resulting in much larger dissolution rates in water compared to vanadates. For e.g., Ke Gu et al. obtained SRO membrane of 4 mm x 4 mm from an SRO (100-300 nm)/SAO (20 nm)//STO heterostructure by dipping it inside water for one day [30]. However, the high reactivity to the atmosphere renders SAO//STO templates unsuitable for further ex-situ film deposition, that includes breaking the vacuum and depositing functional oxides using other deposition chambers or through non-vacuum-based techniques such as chemical solution deposition (CSD).

SVO, on the other hand, is an electrically conductive and optically transparent oxide with a perovskite structure, allowing for easy epitaxial growth of other perovskite oxides on it [31,32]. More importantly, although water soluble, it is more stable under an atmosphere with much lower dissolution rates than that of SAO. In STO/SVO (20 nm)//STO heterostructure, Yoan B et al. reported that it takes five days for SVO dissolution in water, for obtaining a 2.5 mm x 2.5 mm area of free-standing STO [25]. Here we ask the question whether it is possible to design a hybrid water-soluble template combining two seemingly contradictory properties, i.e., stability in the atmosphere and faster dissolution rates in water (comparable to reactive SAO).

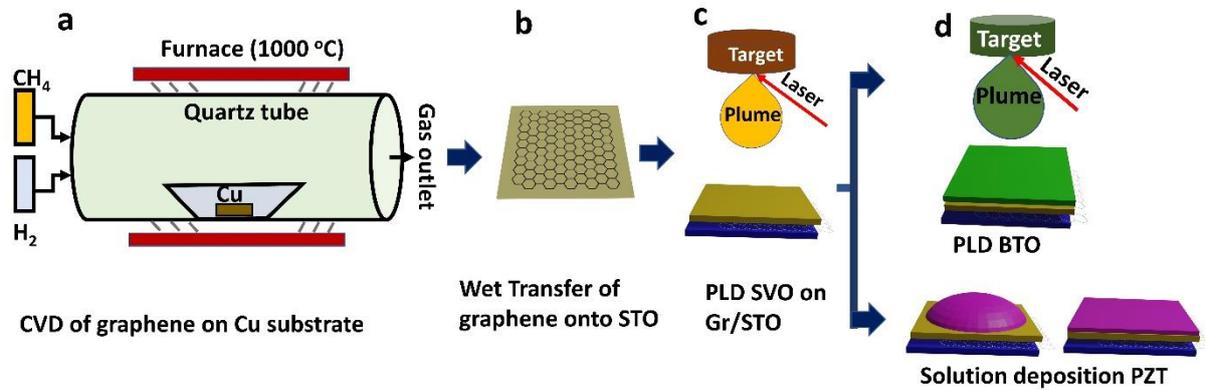

**Figure 1: Schematic representation of the thin film fabrication process.** (a) chemical vapor deposition (CVD) of monolayer graphene on polycrystalline copper sheet, (b) wet transfer of graphene from copper to STO substrate, (c) Pulsed laser deposition (PLD) of SVO on graphene-coated STO substrate (deposition of SVO/Gr//STO template), (d) PLD growth of BTO and chemical solution deposition (CSD) of PZT on the SVO/Gr//STO template.

Remote epitaxy is recently gaining popularity as a layer transfer method that relies on growing the complex oxide on a lattice-matched substrate through a 2D layer at the interface, thereby reducing the substrate-film interaction and enabling top-layer transfer [33–38]. Transfer of oxides such as STO grown via hybrid molecular beam epitaxy (MBE) [34]; $VO_2$, $SrTiO_3$ (STO), $CoFe_2O_4$ (CFO), $Y_3Fe_5O_{12}$ (YIG), $LiNbO_3$ (LNO) by pulsed laser deposition (PLD) [33,39,40], among others, has been demonstrated through remote epitaxy using graphene. In addition, graphene at the interface of the sacrificial layer and substrate enhances the etchant diffusivity compared to the interface without graphene, leading to an increased removal rate of the sacrificial layer [41].

In this study, we utilized a combination of remote epitaxy (with graphene) and a stable sacrificial layer (SVO) to enable large-area transfer of functional oxides at enhanced rates (comparable to the dissolution of atmospherically reactive SAO). Our hybrid template is a heterostructure of SVO/Gr//STO, with graphene grown via large area CVD and transferred onto STO and SVO grown via PLD (Figure 1(a-c)). We clearly demonstrate a 10-fold increase in the dissolution rates of SVO with graphene at the SVO//STO interface. On these hybrid templates, we grew a) epitaxial ferroelectric BTO using PLD without breaking the vacuum, and b) oriented PZT using chemical solution deposition (CSD), which involves exposing the template to atmosphere and liquid solutions (Figure 1(d)). Our hybrid template has superior atmospheric stability and robustness to harsher processing. We transfer these membranes to a Si substrate and demonstrate their ferroelectricity. Additionally, we demonstrate reusability of the STO substrate for further growth of thin film/SVO bilayer.

**Results and Discussion:**

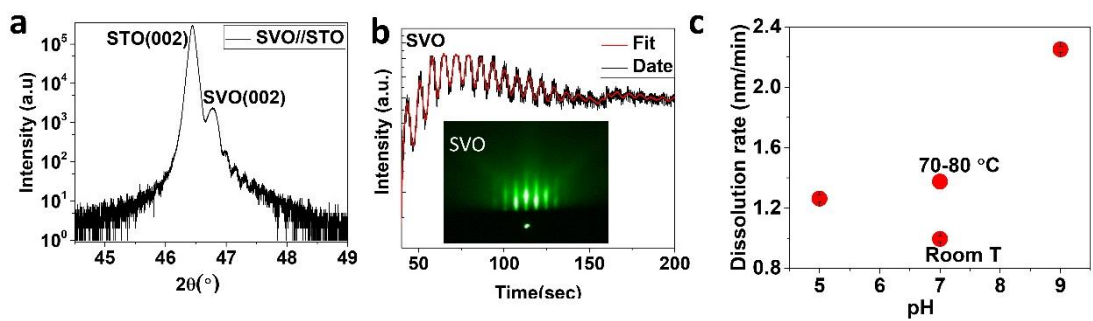

**Figure 2: Water soluble single crystalline SVO on STO substrate.** (a) X-ray diffraction θ-2θ scan zoomed around the (002) Bragg reflection of SVO, (b) Specular spot RHEED oscillation during SVO growth (black: raw data and red: Gaussian fit). Clear oscillations can be seen corresponding to one unit cell growth of SVO. Inset shows the RHEED image of SVO after 1200 pulses. (c) Vertical dissolution rate (in nm/min) of SVO in a different pH and temperature solution.

**Deposition and characterization of water-soluble SVO on STO substrates**

To understand and quantify the water-soluble nature of SVO, we first deposited SVO layers epitaxially directly on the STO substrate using PLD (see methods). Figure 2(a) shows the XRD out of plane θ-2θ scan on the reference sample around STO's (002) Bragg reflection. The (002) Bragg peak of SVO film appears at 46.78°, with the Laue fringes, and the FWHM of symmetric rocking curve for (004) Bragg reflection is $0.08^0$, all revealing excellent crystalline quality of SVO (refer to Figure S1 (a-d) for symmetric and asymmetric scans and samples of various thicknesses, Figure S2 for XPS). Figure 2(b) shows RHEED intensity oscillations of as-grown SVO film, revealing the layer-by-layer growth mode of the film, and AFM shows that the film follows substrate terraces. (See Figure S1(e)). High-resolution TEM (HRTEM) image (Figure S1 f) shows a good interface between SVO and STO (001).

The electrical resistivity value of the SVO layer was measured (by van der Pauw method) as $(2.4\pm0.05) \times 10^{-4}$ ohm.cm, similar to the reported values for PLD-grown metallic $SrRuO_3$ (SRO) and SVO thin films on STO substrates [42,43] (see table TS2). It may be noted that upon the deposition of a BTO layer on the top of SVO, the resistivity value of SVO increases to $(3.3\pm0.04) \times 10^{-4}$ ohm.cm.

The water solubility of the SVO layer grown on STO substrates was studied. The solubility test was conducted by marking both the as-grown SVO on STO and a reference STO sample with a red stencil marker, followed by its immersion in water. The disappearance of the red marks from SVO on the STO sample indicated the water solubility of SVO, while the reference STO sample retained the mark, as shown in Figure S3.

Further experiments were conducted to alter the dissolution rates of SVO in solution by varying pH, temperature, and stirring. A photoresist (PR) was spin-coated onto SVO and was patterned through photolithography, leaving areas of bare SVO and patterned SVO, both of which initially are at the same height. The dissolution rates of bare SVO were quantified by subsequently immersing the samples in deionized (DI) water and measuring the step height using AFM after removing the PR (see Figure S4). The results showed that the SVO dissolves in water at a rate of $1\pm0.02$, $1.3\pm0.02$, and $2.3\pm0.02$ nm/min for pH of 7, 5, and 9, respectively (Figure 2c). Furthermore, increasing the pH 7 DI water temperature enhanced the vertical etch rates of SVO to $1.4\pm0.03$ nm/min, as shown in Figure 2(c). In the following, we use only DI water at pH 7 (and elevated temperatures of 80-90 °C) to avoid damage to the functional oxide layer upon its exposure to acidic/basic solvents.

**Graphene-coated STO substrates:**

To create graphene coated STO, large-area graphene with full coverage was first grown on a copper substrate of 1-inch x 3-inch dimensions in a custom-made chemical vapor deposition system, details of which can be found in ref [44]. Graphene was then transferred onto an STO substrate using a wet transfer process [details in Figure S5 (a-h)]. Raman spectroscopy results for graphene on STO at room temperature are compared with that exposed to 750 °C and $5\times10^{-6}$ mbar inside a vacuum chamber in Figure S5 (i). Given that graphene and STO have similar spectral features, the spectrum of graphene was obtained by subtracting the reference STO spectrum from that of graphene on STO (shown in Figure S5S (j)) [45]. By performing a

Lorentzian fit, the peak positions for the D, and 2G peaks were determined to be 1584 and 2680 cm$^{-1}$, respectively. The presence of the D and 2G peaks in graphene on STO samples treated at 750 °C, and 5x10$^{-6}$ mbar pressure inside the PLD chamber confirms the stability of graphene at these conditions.

Next, hybrid templates of SVO/graphene were synthesized on STO, combining wet transfer of CVD grown graphene and pulsed laser deposited SVO using the same optimized growth conditions of SVO discussed earlier (also see methods).

**Growth of BTO on hybrid templates using PLD:**

BTO was deposited on SVO/Gr//STO template and for comparison also on SVO//STO template, using growth conditions described in the methods. We refer to BTO/SVO//STO sample as a reference sample in this section and BTO/hybrid template as sample A. HAADF-STEM and corresponding EDS analysis show distinct layers and sharp interfaces [Figure S6 (a-f)]. It may be noted that the order of layers in SVO/Gr//STO heterostructure is quite crucial for maintaining single crystallinity and epitaxy of the BTO layer, and exchanging SVO and Graphene layers (Gr/SVO//STO) did not yield epitaxial growth of BTO due to damage to the SVO surface during wet graphene transfer. [See RHEED data in supplementary Figure S7 (a, b, c, d)].

XRD θ-2θ scans of (002) Bragg peaks of BTO layers and their corresponding rocking curves of BTO layers in both the reference stack and sample A with graphene at the interface are compared in Figure 3(a) [also see Figure S8 (a) for a full scan from 20º-110º]. The out-of-plane lattice parameter (c) of BTO reduces from 4.08±.004 Aº in reference sample to 4.05±.006 Aº in sample A with Gr at the interface (note that bulk value of c in BTO is 4.04 Aº [46]), clearly revealing that remote epitaxy through graphene relaxes stress in the BTO layer. Furthermore, the FWHM of the symmetric rocking curves of sample A is 0.8º, larger than 0.5º, measured in the reference sample, suggesting a slight deterioration of the crystal quality, which could be a consequence of defects such as folds and ripples in the graphene caused by the wet-transfer process. The epitaxial nature of the BTO films is confirmed by studying the XRD φ-scan of the (103) family of planes [Figure. S8 (b)]. The results show a cube-on-cube heteroepitaxy. The BTO RMS surface roughness is estimated through atomic force microscopy as 0.4 nm [Figure S8 (c)].

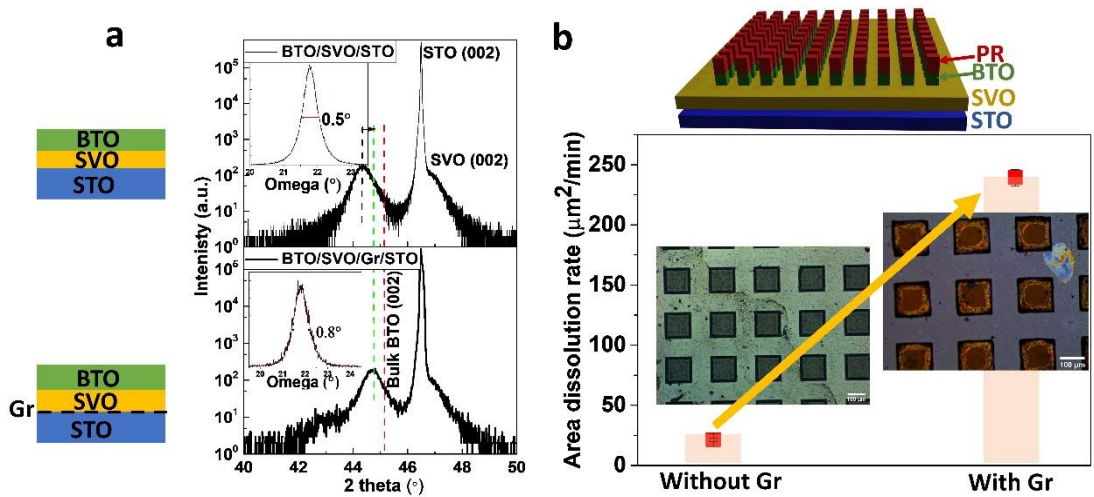

**Figure 3: PLD BTO/SVO//STO (reference sample) and BTO/SVO/Gr//STO (sample A) X-ray diffraction (XRD) and dissolution rates.** (a) XRD θ-2θ scan zoomed around the (002) BTO Bragg reflection [inset showing Omega scan of (002) Bragg reflection of BTO]. (b) Plot illustrating the average dissolution rates represented in μm$^2$/min for reference sample and sample A. The inset displays images of partially dissolved SVO in a square patterned BTO/SVO/Gr//STO (top cartoon showcasing it).

**Graphene-assisted dissolution of SVO:**

To determine the dissolution rates of SVO and the effect of graphene layer, the BTO layer in both the reference sample and sample A was selectively exposed in square patterns and etched following (shown in Figure 3(b) top cartoon) standard photolithography process. These squares were subsequently etched using buffered hydrofluoric acid (BHF) solution. The entire stack was then immersed in DI water (resistivity 17.8-18 Mohm.cm) at temperatures ranging from 70-80 °C, and optical images were taken at regular intervals. The contrast observed at the edges of the square from the optical images confirms the underlying SVO dissolution front (see Figure S9 for more detail). The results of the study indicate that the inclusion of graphene increases the dissolution rates of SVO from 21.35±1.5 μm$^2$/min in the reference sample (without graphene) to 239.35±6.82 μm$^2$/min for sample A (with graphene at the interface), as illustrated in Figure 3(b) (also see Figure S10). Our observation of a massive increase (10 times) in dissolution rates in the presence of graphene-lined channels is consistent with earlier reports that showed an enhancement in diffusivity of the etchant medium (e.g., water) through graphene channels [41]. This is argued to be a consequence of a new surface diffusion mechanism in graphene-like layered materials, whereby nanodroplets of water surf through propagating ripples [47].

**Fabrication of Free-standing (FS) BTO film and transfer process:**

Next, we employed two different methods to transfer BTO functional oxide layer onto Si substrates:

a. Stamping method

b. Floating method

**a. Stamping method:**

Here, we use a top handling that binds to the BTO layer [25]. Both Polydimethylsiloxane (PDMS) coated with polypropylene carbonate (PPC) and thermal release tape (TRT) were explored as the handling layers. This layer effectively seals the BTO, making it challenging to dissolve SVO by merely submerging the as-grown heterostructure in water. To overcome this, BTO squares (150 μm x 150 μm) were patterned using photolithography, and the rest of it was wet etched using buffered hydrofluoric acid (BHF). This enabled the creation of many more water channels for SVO dissolution (Figure 4a).

Then sample A was dipped in DI water at 70-80 °C for SVO dissolution. The dissolution front of the underlying SVO layer was monitored continuously by observing it under an optical microscope at regular intervals. Before the SVO was completely dissolved, the sample was removed from the solution, dried, and attached to a mechanical handling layer. The sample was then placed back in the water to remove the remaining SVO, and the BTO membrane was mechanically separated onto the TRT or PDMS. The TRT or PDMS with the free-standing BTO was then stamped onto the target substrate, and the BTO layer was released by heating it beyond 100 °C. A clean (free from organic residue) and lower crack density transfer is achieved using PPC/PDMS handling layer compared to TRT while adopting the stamping route (refer to Figure S11). Note that in the reference sample (without graphene-lined channels), this technique resulted in a transfer of BTO squares with large crack densities (see Figure S12).

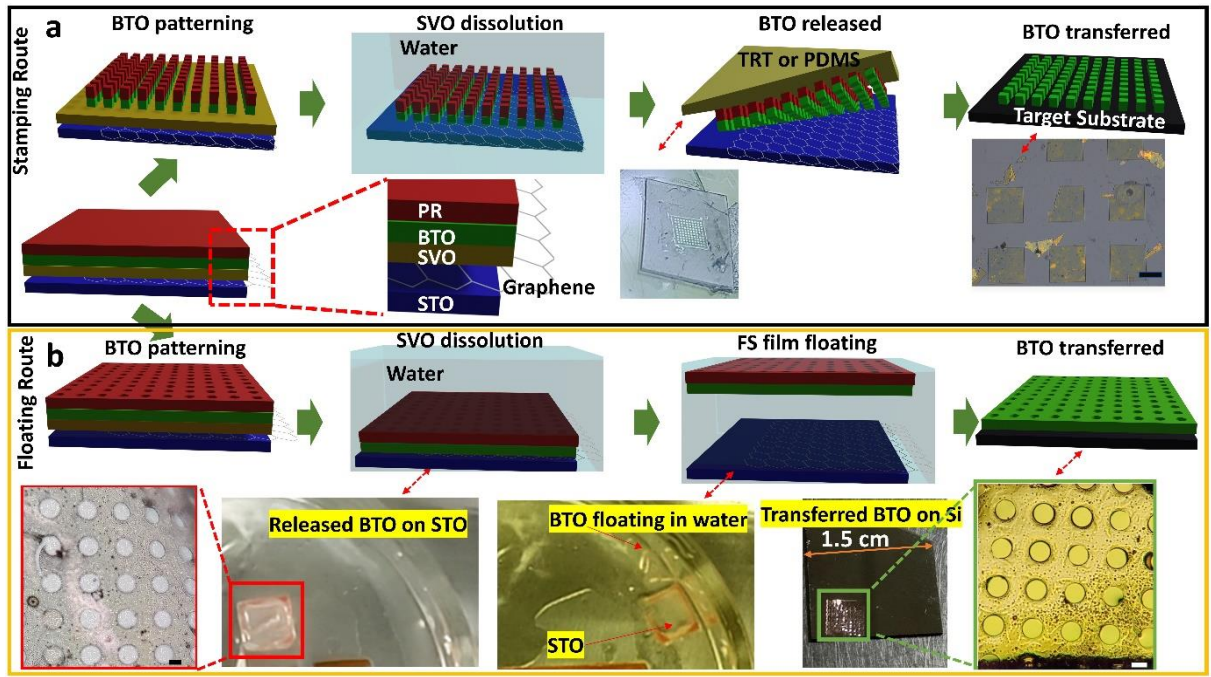

**Figure 4: Schematic of obtaining FS film onto target substrate of Si.** (a) Stamping route: Sample A with graphene at the interface patterned into squares through standard photolithography followed by wet etching BTO by BHF, dissolution of SVO, and BTO transferred using TRT or PDMS (corresponding optical image are shown with dashed red arrows). (b) Floating method: Sample A patterned with circular grids, SVO dissolved in water, then FS film floated and scooped with Pt-Si (corresponding optical images are shown in the bottom row).

b. **Floating method:**

To enable large-area transfer, residual stresses in the membrane need to be released first. To this end, we patterned circular grids (of diameter 140 μm) using photolithography and wet-etched BTO along these patterns [Figure 4(b)]. Square grids were avoided as they resulted in cracks forming at the corners (stress concentrators) during the dissolution of the underlying SVO layer, as shown in Figure S13 (a, b). After dissolving the SVO layer, the sample was removed from water, dried, and gradually immersed at an angle, utilizing the capillary action of water to release the BTO layer. The BTO membrane (area ~3.5 mm x 3.5 mm) with photoresist on top was then floated and scooped with the target substrate in one go. Video S1 demonstrates the release of the BTO layer in water and its subsequent scooping onto a target Si substrate, with corresponding images shown in Figure 4b (bottom row). One benefit of the floating method is that adhesion between the free-standing film and the target substrate is primarily due to water drying from the interface and surface tension, enabling the use of virtually any target substrate and a large area membrane transfer. In contrast, the stamping method is restricted by the constraint that interface energy between the free-standing film and the target substrate should be less than the film and the polymeric handling layer, making it critical to select a suitable target substrate for the transfer process.

It is worth noting that floating a large area is not practically feasible without the graphene-lined nanochannels (i.e., in the reference sample).

**Structure and property characterization of BTO free-standing films:**

XRD θ-2θ scan of FS BTO transferred onto a Pt-coated Si substrate shows (00l) Bragg reflections BTO (Figure S14 (a), confirming that BTO retains its well-oriented crystalline structure even after the transfer process. Figure 5(a) depicts the 2θ position from the (002) Bragg reflection for BTO, both as-grown and after the transfer, revealing a relaxation in the out-of-plane lattice parameter from 4.05±.006 A° to 4.02±.003 A° once released (c in bulk=4.04 A°). The asymmetric θ-2θ scan in Figure S14 (b, c) [showing (103) Bragg reflection for the clamped film and in-plane scan for FS film] indicates an increase in the in-plane lattice parameter from 3.94 A° to 3.96 A° due to the release of an in-plane compressive strain of 0.5%. Furthermore, the symmetric omega scan of the (002) peak in Figure S14 (d) shows that the FWHM of the rocking curve in the FS film is comparable to that of the clamped BTO film (0.8°). The transferred FS film onto Pt/Si substrate exhibits sub-nm surface roughness, as demonstrated in Figure S15 (a), while the AFM step height of the FS membrane shows a uniform film thickness of 180 nm. (Figure S15 (b)).

Piezo-response force microscopy (PFM) was used for a quick check of the ferroelectric properties of the BTO membrane. A positive bias of 7 V was applied in an area of 5 μm x 5 μm, followed by -7 V at the central 2.5 μm x 2.5 μm area. After that, a small AC voltage of 0.7 V is applied onto the tip to read the contrast between differently poled regions. PFM phase and amplitude map are shown in Figure 5(b) and (c), respectively. The phase map shows a 180° contrast across the oppositely poled regions, indicating that the ferroelectric domains were written and, thus, evidencing ferroelectricity in these films.

**Growth of PZT on the hybrid template using solution deposition technique and its structure-property studies:**

The relative stability of SVO in the atmosphere compared to other standard water-soluble layers, such as SAO, renders our hybrid templates versatile ex-situ templates. We demonstrate this versatility by growing PZT on our hybrid templates through chemical solution deposition (CSD, see methods) and its subsequent transfer (Figure S16). It is important to note that our templates are not only stable in the atmosphere but also upon long exposure to the organic liquid solutions involved in the CSD technique (methods).

The highly oriented nature of the as-grown PZT through CSD and PZT membrane transferred onto PDMS was confirmed using an out-of-plane theta-2theta XRD scan, as shown in Figure 5(a). PFM writing (of PZT membrane on Pt-Si) again shows a 180º phase difference between different domains, indicating the ferroelectric nature of PZT films too.

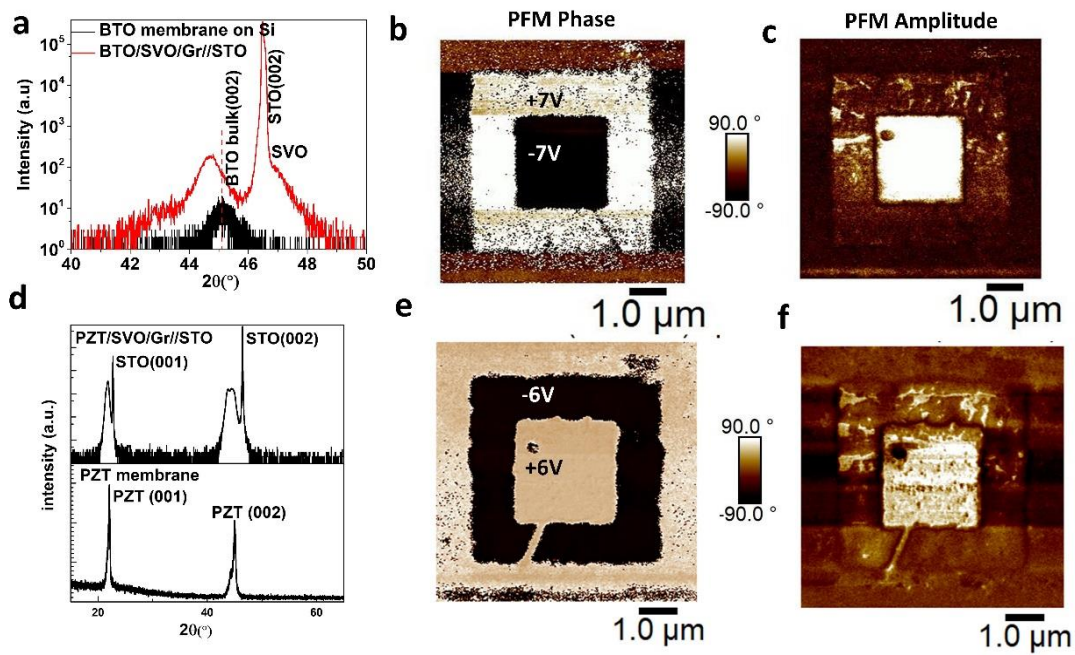

**Figure 5: Characterization of the FS membrane.** (a) XRD out-of-plane θ-2θ scan near (002) Bragg reflection of BTO in sample A (BTO/SVO/Gr//STO) and FS BTO transferred onto Pt-Si. PFM (b) Phase and (c) amplitude image of BTO. Outer 5 μm x 5 μm area with 7 V and inner 2.5 μm x 2.5 μm with -7 V DC. (d) XRD out-of-plane θ-2θ scan of PZT/SVO/Gr//STO and FS PZT on PDMS. PFM (e) phase and (f) amplitude image of FS PZT on Pt-Si. Outer 5 μm x 5 μm area with -6 V and inner 2.5 μm x 2.5 μm with 6 V DC.

**Reusability of the STO substrate:**

Finally, we demonstrate that our hybrid template approach allows for efficient recyclability of the substrate, which is a crucial cost-saving factor in oxide substrates such as STO. To this end, after the first transfer of FS BTO onto Si from both the hybrid template and the reference template, the parent STO substrates (referred to as Substrate A and Substrate R, respectively) were cleaned ultrasonically and reused for the growth of SVO/BTO (Figure 6 a, also see Figure S17).

Figure 6(b, c) shows that the surface roughness of STO after the first transfer process for the graphene-coated STO parent substrate is 1.1 nm, while without the graphene-coated STO parent substrate, it is 1.9 nm. Furthermore, the FWHM of the rocking curve for the symmetric (002) Bragg reflection for regrown BTO on substrate A is 0.9°, which is better than that on the reference substrate (R) 1.1° [Figure 6(d)], clearly showing that graphene layer enhances substrate's suitability for subsequent reuse.

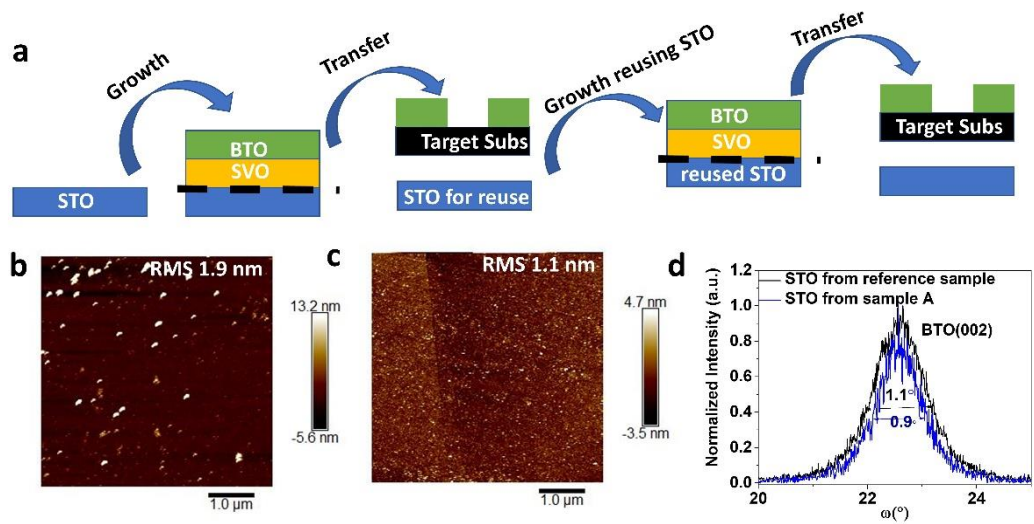

**Figure 6: STO substrate reusability with and without graphene at the interface.** (a) Schematic representation. AFM surface roughness of STO after first transfer for (b) reference sample without graphene and (c) sample A with graphene at the interface. (d) Omega scan of (002) Bragg reflection of BTO in BTO/SVO grown on reused STO from reference sample (without Gr) and sample A (with graphene at the interface).

**Conclusion:**

In this work, we demonstrate the efficiency of a hybrid template of water-soluble SVO (PLD grown)/graphene (large area CVD growth and transferred)//STO for complex oxide growth and transfer. This template is atmospherically stable yet shows comparable dissolution rates with that of more reactive and commonly used water-soluble layers such as SAO. In order to show the robustness of these templates, we grew BTO using PLD and PZT using chemical solution deposition, heterogeneously transferred large areas of these membranes deterministically to Si substrate, and demonstrated their ferroelectric properties. We show that the substrate-induced strains on these membranes are relaxed upon their release. Importantly, we show that our hybrid template and subsequent transfer process render the parent substrate reusable for further depositions. Fundamentally, we show that templates created by a combination of remote epitaxy and stable water-soluble layers can be used in versatile deposition processes. Such a template design is crucial in the heterogeneous integration of various complex oxides onto Si and flexible substrates for micro and flexible electronics.

**Methods:**

**Growth of SVO and BTO thin film:** STO substrates (Testbourne, Germany) were processed to have a single-terminated $TiO_2$ surface with steps and terraces. This was achieved by solvent-cleaning the substrates in an ultrasonicator bath for 5 minutes, followed by removal of the SrO termination by dipping into a buffered hydrofluoric (BHF) acid solution for 20 seconds, and then annealing at 950 °C in an oxygen ambient for 1 hour.

The SVO PLD target was prepared by mixing a stoichiometric ratio of $SrCO_3$ and $VO_2$ powders in a mortar and pestle for 2 hours, calcining at 850 °C for 5 hours, grinding for 2 hours, and sintering green pellets at 950 °C for 5 hours. Similarly, the BTO target was fabricated using nano powders of BTO, pelletized, and sintered at 1350 °C for 12 hours.

SVO and BTO were subsequently grown on the Gr/STO and STO substrate inside a reflection high-energy electron diffraction (RHEED)-assisted PLD chamber using a KrF excimer laser with a wavelength of 248 nm. An SVO polycrystalline target was ablated using a laser fluence of 1.4 $J/cm^2$, a laser frequency of 1 Hz under a vacuum of $1 \times 10^{-6}$ mbar, and a substrate temperature of 750°C. Similarly, the BTO target was ablated with a laser fluence of 1.6 $J/cm^2$, a laser frequency of 1 Hz, and a substrate temperature of 800 °C. The BTO thin film was grown in a two-step pressure regime where the initial few nm of BTO were grown under a vacuum of $5 \times 10^{-6}$ mbar, and the oxygen pressure was ramped to $5 \times 10^{-2}$ mbar for the rest of the growth. A constant substrate-to-target distance of 5 cm was maintained for all depositions.

**CVD graphene growth and transfer:**

Graphene was grown on both sides of a polycrystalline Cu substrate in a specially designed chemical vapor deposition (CVD) reactor [Figure 1 (a)].

Initially, the furnace was gradually heated over 50 minutes until reaching a growth temperature of 1040 °C in Ar ambient using a flow rate of 150 SCCM. The foil was then annealed at the same temperature for 60 minutes, using a hydrogen flow rate of 150 SCCM. Throughout the graphene growth stage (30 min), methane gas was supplied at 0.75 SCCM, while hydrogen gas was supplied at 150 SCCM (with a supersaturation ΔG of 72.10 kJ/mol). After that, post-growth annealing for 5 min using a methane flow rate of 3 SCCM (ΔG of 83.7 kJ/mol) and finally, cooling down in hydrogen ambient using a flow rate of 150 SCCM. A constant reactor pressure of 1 torr was maintained throughout the ramp-up, annealing, graphene growth, post-growth annealing, and cooldown.

Polymethyl methacrylate (PMMA) was spin-coated at a speed of 2000 revolutions per minute (rpm) to transfer graphene from Cu foil. The graphene layer on the bottom side was subsequently eliminated using a 1:1 ratio of $HNO_3$ to deionized (DI) water. To etch the copper foil, an ammonium persulphate solution was utilized. The remaining graphene and the PMMA layer were transferred into DI water to dissolve any salt residue. The purified graphene was then scooped with STO substrate (of lateral dimensions 5mm x 5mm) and dried. Finally, the PMMA layer was removed by immersing it in acetone overnight, followed by rinsing it with fresh acetone and isopropyl alcohol to eliminate any PMMA residue. For a more detailed, step-by-step procedure, please refer to Figure S5.

**Chemical solution deposition of PZT:**

The chemical solution deposition method was utilized to prepare $Pb(Zr_{0.52}Ti_{0.48})O_3$ (PZT) thin films with a stoichiometric composition [48,49]. The starting materials were Lead(II) acetate Trihydrate, Zirconium(IV) n-propoxide, and Titanium(IV) iso-propoxide, while 2-methoxy ethanol, glacial acetic acid, and n-propanol were used as solvents, and formamide and acetylacetonate as stabilizing agents. Additionally, Polyvinylpyrrolidone (PVP) was used as a modifier to obtain a thicker, crack-free film.

To prepare the thin film, a 0.1M solution of Pb precursor was made by dissolving Lead acetates in glacial acetic acid. Pre-dissolved Zr and Ti propoxides in 2-methoxy ethanol were added to the Pb precursor solution while stirring. Then, distilled water was added for acid hydrolysis, followed by the addition of formamide and acetylacetonate, and finally, PVP. The solution was stirred for one hour, filtered using 100 μm filter paper, and spin-coated at 4000 rpm for 30 sec on the PLD SVO/Gr//STO template, followed by pyrolysis at 400 ºC for 5 min. The deposition cycles were repeated until the desired 150 nm thick film thickness was obtained. Finally, post-annealing was carried out at 700 ºC for 1 hr in an open-air atmosphere furnace.

**XRD, AFM, Raman, TEM, XPS**

For X-ray diffraction (XRD), a four-circle XRD instrument (Rigaku smart lab) with a Cu Kα source was utilized. The film's surface morphology was studied using atomic force microscopy (AFM) with a Bruker Icon Dimension system. Poling during piezo-force microscopy (PFM) measurements were conducted using Pt-Ir coated conducting AFM tips. Raman spectroscopic measurements with a LabRAM HR spectrometer were employed to investigate the presence of the graphene layer using a 532 nm laser source. For cross-sectional transmission electron microscopy (TEM), a lamella was prepared using a focused ion beam (FIB) and studied using TEM (Titan Themis from FEI) at 300kV operating voltage.

The film's chemical analysis was carried out using X-ray photoelectron spectroscopy (XPS) with a monochromatic Aluminium source in a Kratos axis ultra XPS system equipped with a magnetic immersion lens, a charge neutralization system, and a spherical mirror analyser. The spot size used was around 15 microns. The survey and high-resolution spectra were collected using 1 eV and 0.1 eV resolution, respectively.

## ASSOCIATED CONTENT

**Supporting Information**.

The following files are available free of charge.

   XRD, TEM, XPS, RHEED, and other characterization data (PDF)

   Free-standing BTO transferred through the floating method (MP4 Video)

## AUTHOR INFORMATION

**Corresponding Author**

*Srinivasan Raghavan, Email: sraghavan@iisc.ac.in

*Pavan Nukala, Email: pnukala@iisc.ac.in

*Asraful Haque, Email: asrafulhaque@iisc.ac.in

**Present Addresses**

*Center for Nanoscience and Engineering, Indian Institute of Science, Bengaluru, India, 560012

**Notes**

The authors declare no competing financial interest.

**ACKNOWLEDGMENT**

This work was partly carried out at the Micro and Nano Characterization Facility (MNCF) and National Nanofabrication Center (NNfC) located at CeNSE, IISc Bengaluru, funded under Grant DST/NM/NNetRA/2018(G)-IISc, NIEIN, and Ministry of Human Resource and Development, Government of India and benefitted from all the help and support from the staff. P.N. acknowledges Start-up grant from IISc, Infosys Young Researcher award, and DST-starting research grant SRG/2021/000285. AH acknowledges Rama Satya Sandilaya Ventrapragada for helping with PFM measurements. AH also acknowledges Mrityunjay Pandey and Professor U. Chandni for helping in PPC based deterministic transfer process.

**ABBREVIATIONS**

PLD Pulsed laser deposition, CSD chemical solution deposition, TRT thermal release tape, PDMS Polydimethylsiloxane, PPC poly(propylene) carbonate.

# Supporting Information

# Robust atmospherically stable hybrid SrVO$_3$/Graphene//SrTiO$_3$ template for fast and facile large-area transfer of complex oxides onto Si

Asraful Haque, Suman Kumar Mandal, Antony Jeyaseelan, Sandeep Vura, Pavan Nukala, Srinivasan Raghavan.

Center for Nanoscience and Engineering, Indian Institute of Science, Bangalore

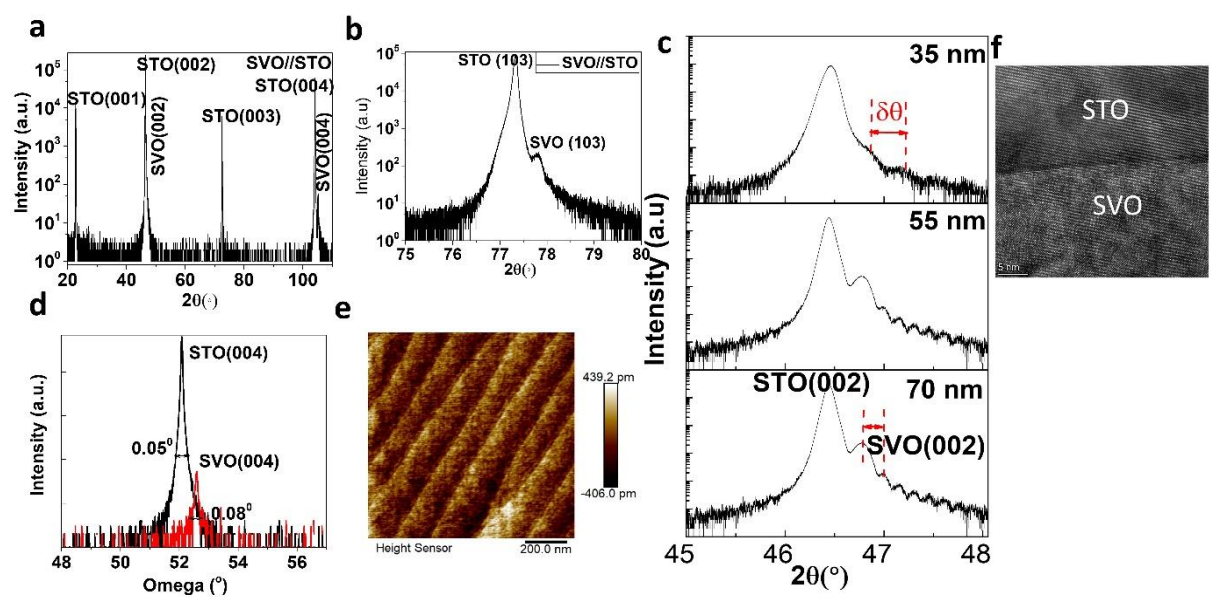

**Figure S1: SVO on STO characterization.** Out of plane θ-2θ XRD pattern of SVO//STO bilayer from (a) 20-110º, (b) near asymmetric (103) Bragg reflection of SVO, (c) near symmetric (002) Bragg reflection of SVO for different thickness of 35nm, 55 nm, and 70 nm. (d) Symmetric omega scan for (004) Bragg reflection of STO and SVO with FWHM of $0.05^0$ and $0.08^0$, respectively. (e) AFM image of the surface morphology. (f) HRTEM image showing the interface of SVO//STO.

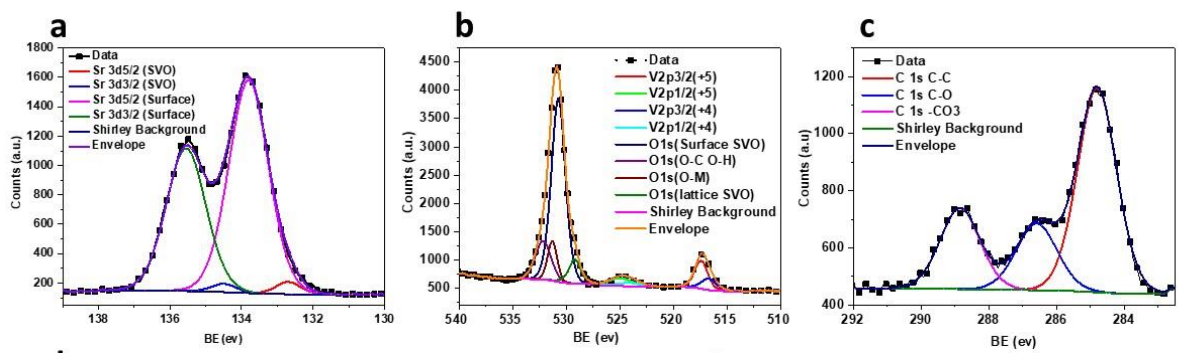

**Figure S2: Chemical composition analysis as-grown SVO on STO.** High-resolution XPS spectra of (a) Sr 3d, (b) V 2p-O1s, (c) C 1s.

*XPS analysis of SVO:*

The high-resolution spectra of the elements Sr 3d, V 2p, and O 1s have been collected, as shown in Figure S2. All peaks are corrected using an adventitious C peak at 284.8 eV and fitted using a Lorentzian-Gaussian function (in CASAXPS software) and Shirley background to remove the secondary electron background [1]. Fitting parameters like BE, FWHM, and associated peak contribution are tabulated in Table TS1.

*Fitting procedure for O1s-V2p*: Four contributions were attributed to the O1s peak. The first two contributions are lower BE of 529.22 eV and 530.73 eV, with FWHM of 1.2 ev and 1.5 eV. They are attributed to SVO lattice, more particularly to O 1s from the SVO lattice site and O 1s from surface SVO, which is an oxygen-rich surface layer. The third contributions are from the oxygen atom linked to metal/metal hydroxide, denoted as O 1s(O-M) at 531.29 eV with FWHM 1.09 eV. The fourth contribution is oxygen atoms connected to organic surface contamination or physio-absorbed species represented as O 1s(O-C, O-H) at 532.15 eV with an FWHM of 1.7 eV.

*Fitting of V2p region*: A perfect SrVO$_3$ crystal V should have a +4 oxidation state. However, multiple oxidation states of V have been reported in the literature for SVO [2]. Therefore, V 2p$_{2/3}$ can be deconvoluted into two peaks at 517.42 eV and 516.71 eV for V$^{+5}$ and V$^{+4}$ oxidation states, respectively. The FWHM value for these peaks increases from 1.43 eV to 1.66 eV. This is because the +5 state does not have unpaired electrons in the d orbital, unlike the +4 oxidation state of V [3].

Following the spin-orbital splitting, V $2p_{1/2}$ peaks can be constructed with ΔE of 7.5 eV and 7.28 eV for +5 and +4 oxidation state of V and maintaining an area ratio of 1:2. The FWHM of V $2p_{1/2}$ (+5) and V $2p_{1/2}$ (+4) are 2.63 eV and 2.70 eV, respectively. The difference in the FWHM of $2p_{3/2}$ and $2p_{1/2}$ peak is because of the Coster-Kroning effect, similar to what has been observed for Ti 2p level [4].

***Fitting of Sr3d region***: For Sr $3d_{5/2}$, two different peaks at 132.94 eV and 135.82 eV have been identified corresponding to Sr bonded to lattice denoted as "Sr $3d_{5/2}$(SVO)," and the other Sr bonded to oxygen or hydroxide indicated as "Sr $3d_{5/2}$(Sr-O)". The FWHM of these peaks is 0.8 eV and 1.3 eV, respectively. Following the spin-orbital splitting, Sr $3d_{3/2}$ peaks can be constructed with ΔE of 1.8 eV and an area ratio of 0.67. The corresponding FWHM is 0.8 eV and 1.3 eV for SVO lattice and surface, respectively.

**Table TS1:** Peak assignment to different chemical elements at the surface of SVO and their corresponding fitting parameters, Binding energy (BE) in ev, Full width at half maximum (FWHM). The corresponding fitting is demonstrated in Figure S2.

| Name | Position (eV) | FWHM (eV) | Doublet separation | % area |
|---|---|---|---|---|
| V $2p_{3/2}$ (+5) | 517.42 | 1.43 | 7.5 | 27.49 |
| V $2p_{1/2}$ (+5) | 524.92 | 2.63 | | 40.24 |
| V $2p_{3/2}$ (+4) | 516.71 | 1.66 | 7.28 | 13.09 |
| V $2p_{1/2}$ (+4) | 523.99 | 2.70 | | 19.17 |
| O 1s (lattice SVO) | 529.21 | 1.2 | | 7.84 |
| O1s (surfaceSVO) | 530.73 | 1.5 | | 67 |
| O 1s (O-M) | 531.29 | 1.03 | | 9.62 |
| O 1s (O-C, O-H) | 532.15 | 1.7 | | 15.53 |
| Sr $3d_{5/2}$ SVO | 132.70 | 0.8 | 1.8 | 1.79 |
| Sr $3d_{3/2}$ SVO | 134.50 | 0.8 | | 1.74 |
| Sr $3d_{5/2}$ (Surface) | 133.80 | 1.3 | 1.75 | 48.94 |
| Sr $3d_{3/2}$ (Surface) | 135.55 | 1.3 | | 47.53 |
| C 1s C-C | 284.8 | 1.36 | | 57.39 |
| C 1s C-O | 286.58 | 1.41 | | 19.46 |
| C 1s -$CO_3$ | 288.85 | 1.38 | | 23.15 |

**Table TS2:** Comparison of electrical resistivity of bottom electrodes from the literature

| Conducting electrode | Resistivity (10⁻³ ohm.cm) | Substrate | Growth technique | Ref |
|---|---|---|---|---|
| SVO | 0.24 | STO | PLD | This work |
| SVO in BTO/SVO/STO | 0.33 | STO | PLD | This work |
| SVO | 0.22 | LSAT | Sputtering | |
| | 0.03 | LSAT | Hyb MBE | 5 |
| | 0.12 | LSAT | PLD | 6 |
| | 0.045 | STO | PLD | 7 |
| | 0.117 | LSAT | PED | 8 |
| | 0.11 | LAO | PLD | 9 |
| | 0.28 | STO | PLD | 10 |
| | 0.34 | Si | E-beam | 11 |
| | | | Laser MBE | 12 |
| | 0.05 | STO | | |
| | 0.24 | STO | PLD | 13 |
| | 0.12 | LAO | PLD | 13 |
| | 0.12 | LSAT | PLD | 13 |
| SRO | 0.2 | Miscut STO | PLD | 14 |
| | 0.22 | Miscut STO | | |
| | 0.23 | Regular STO | | |
| | 0.28 | LAO | | |

|        |       | 0.42  | YSZ |          |
|--------|-------|-------|-----|----------|
|        |       |       |     | 15       |
|        |       |       |     | reactive |
| LSMO   |       | 0.61  | STO | MBE      |
|        |       | 2.9   | LaSrGaO$_4$ | |
|        |       | 16.6  | LAO |          |
|        |       | 82.8  | DyScO$_3$ (DSO) | |
|        |       |       |     | 16       |
| La$_{0.5}$Sr$_{0.5}$CoO$_3$ | | | STO, | |
| (LSCO) | 0.13-0.2 |    | MgO | PLD      |
|        |       | 0.9   | bulk |         |

SVO: SrVO$_3$, BTO: BaTiO$_3$, STO: SrTiO$_3$, LSAT: (LaAlO$_3$)$_{0.3}$(SrAl$_{0.5}$Ta$_{0.5}$O$_3$)$_{0.7}$, LAO: LaAlO$_3$, SRO: SrRuO$_3$, LSMO: La$_{0.67}$Sr$_{0.33}$MnO$_3$ , LSCO: La$_{0.5}$Sr$_{0.5}$CoO$_3$ , YSZ: Yttria stabilized Zirconia, LSGO: LaSrGaO$_4$, DSO: DyScO$_3$, PLD: Pulsed laser deposition, PED: Pulsed electron deposition, MBE: Molecular beam epitaxy, R-MBE: Reactive MBE, H-MBE: Hybrid MBE, E-beam: Electron beam evaporation.

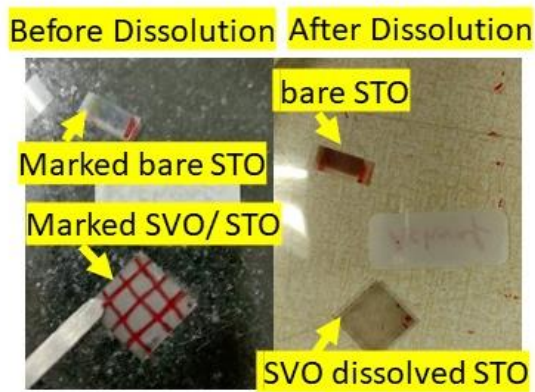

**Figure S3: Dissolution study of the SVO layer.** On the left side: stencil red marked bare STO and SVO//STO samples before dipping them inside DI water. On the right side: after dissolution, marks were retained on bare STO but not the other, confirming SVO is water soluble.

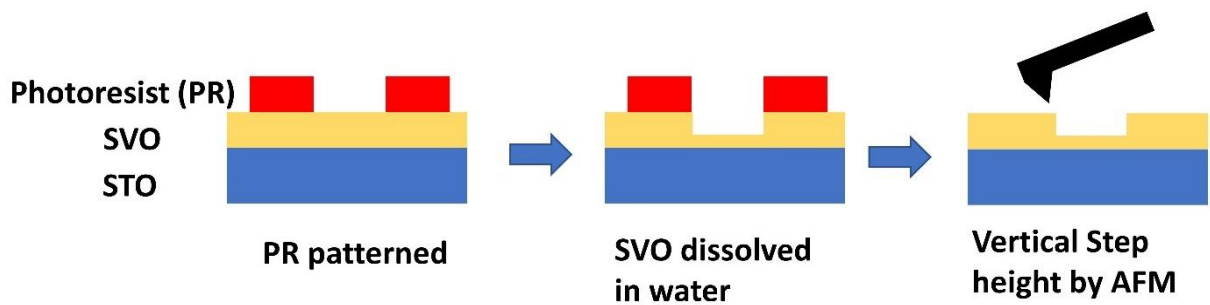

**Figure S4:** Schematic representation of obtaining step for measuring the vertical dissolution rate of SVO using AFM.

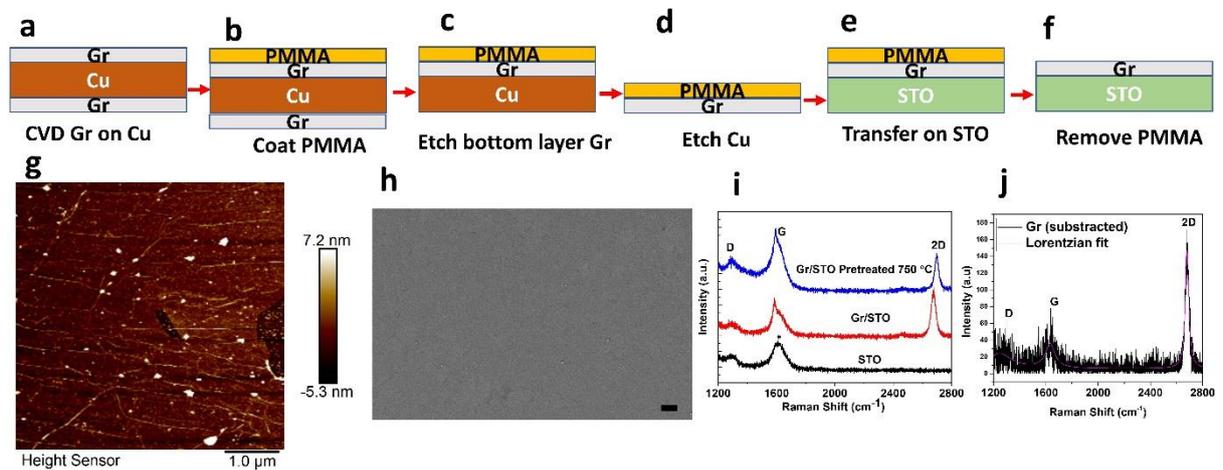

**Figure S5: Characterization of transferred and pre-treated (at 750 °C and 5x10⁻⁶ mbr) graphene.** Schematic of wet transfer of graphene: (a) as grown CVD graphene on Cu substrate, (b) graphene etched (in dilute $HNO_3$ solution for around 10 seconds) from the back side of Cu, (c) top layer graphene on Cu substrate was spin-coated with PMMA for mechanical support, (d) Cu substrate is etched in 0.2M ammonium persulphate $(NH_4)_2S_2O_8$ solution (APS), and then PMMA/Gr stack is floated in DI water to remove the APS residue, (e) thereafter, Gr/PMMA layer is scooped with STO substrate and dried at room temperature, (f) PMMA is removed by dipping inside acetone for overnight. (g) AFM image surface morphology and (h) SEM micrograph of the Gr/STO substrate (scale bar 10 μm). The AFM micrograph shows PMMA residues on the as-transferred Gr on STO through the wet method. (i) Raman spectroscopy collected after pre-treatment of Gr/STO stack. Pre-treatment was done at 750 °C under a 5x10⁻⁶ mbar pressure inside the PLD chamber. Both G and 2D peaks indicate the presence of Gr on STO even after the pre-treatment process. Graphene shares common spectral features with STO in the spectral range of 1400-1800 cm$^{-1}$; therefore, subtracted spectra of Gr//STO from that of STO give the signature of stand-alone Gr peaks as shown in (j).

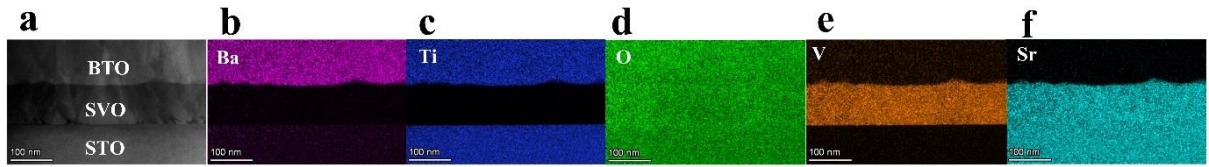

**Figure S6:** (a) Cross-sectional HAADF-STEM image of BTO/SVO//STO heterostructure. (b-f) Chemical composition mapping across BTO/SVO//STO.

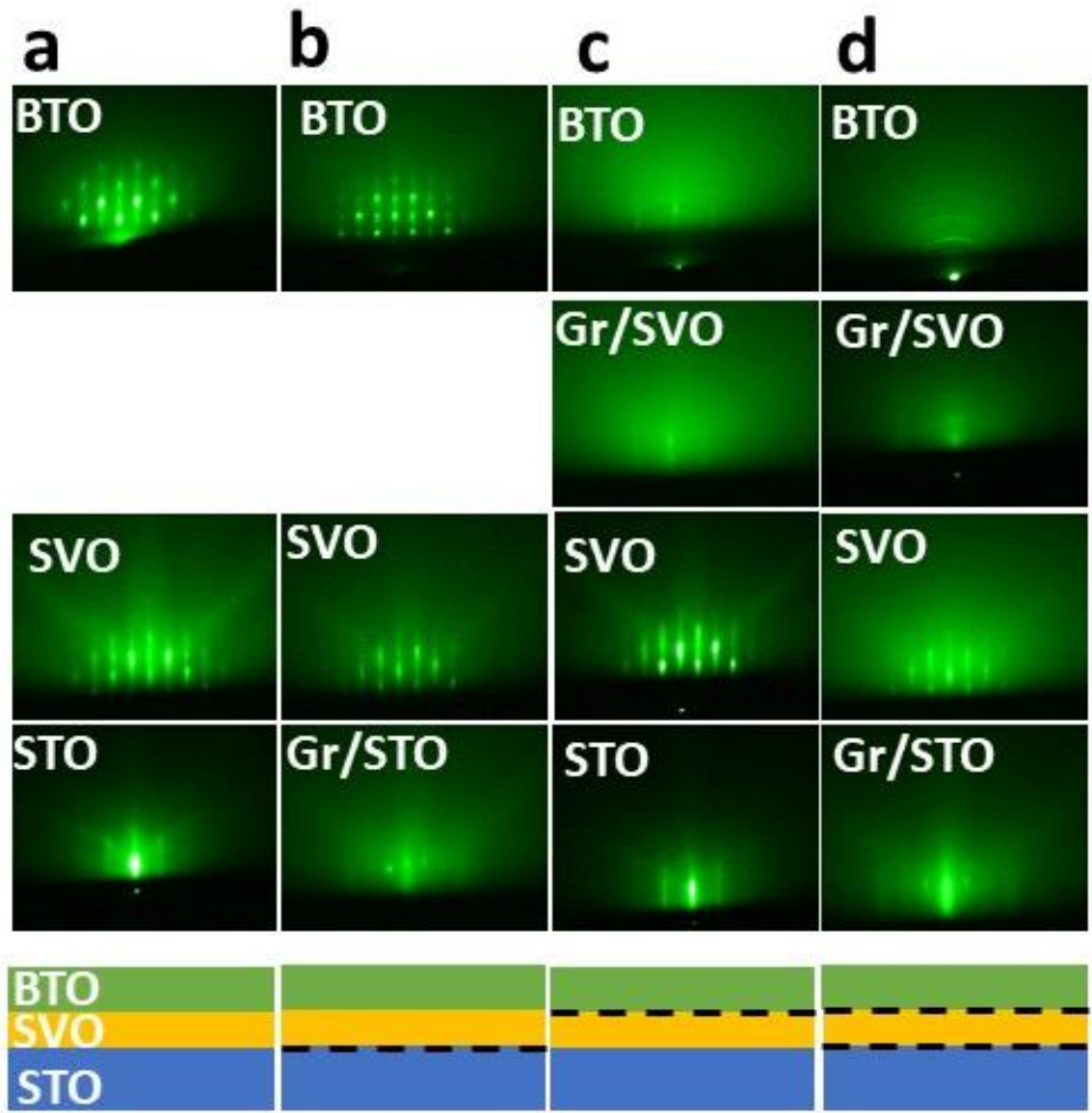

**Figure S7:** RHEED pattern of STO substrate, SVO, and BTO surfaces in (a) reference sample (BTO/SVO//STO), (b) sample A (BTO/SVO/Gr//STO), (c) BTO/Gr/SVO//STO and (d) BTO/Gr/SVO/Gr//STO heterostructure. The vertical panel indicates the RHEED pattern of the surface of the STO substrate (with and without graphene, indicated as Gr/STO and STO, respectively), SVO layer after 1200 pulses, post-graphene transfer onto the SVO surface, and BTO after 1800 pulses, respectively. After the wet transfer of graphene, the SVO surface RHEED pattern is provided in the third row from the bottom. Since we have adopted a wet method for graphene transfer, the transfer process deteriorates the top surface of SVO (shown in the heterostructures (c) and (d)), ultimately hindering the epitaxial growth of BTO.

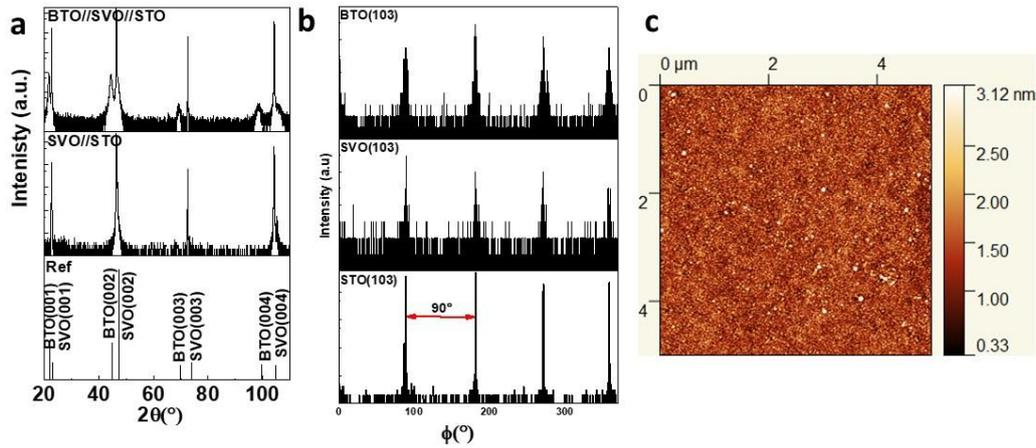

**Figure S8:** (a) XRD symmetric θ-2θ full scan in the range of 20-110° of BTO/SVO//STO and SVO//STO heterostructure, showing (00l) family of planes, (b) Phi scan of asymmetric (103) Bragg reflection of BTO, SVO, and STO shows peaks are at $90^0$ apart representing four-fold symmetry (c) AFM image (5μmx5μm) of the BTO surface after growth showing sub-nm surface roughness.

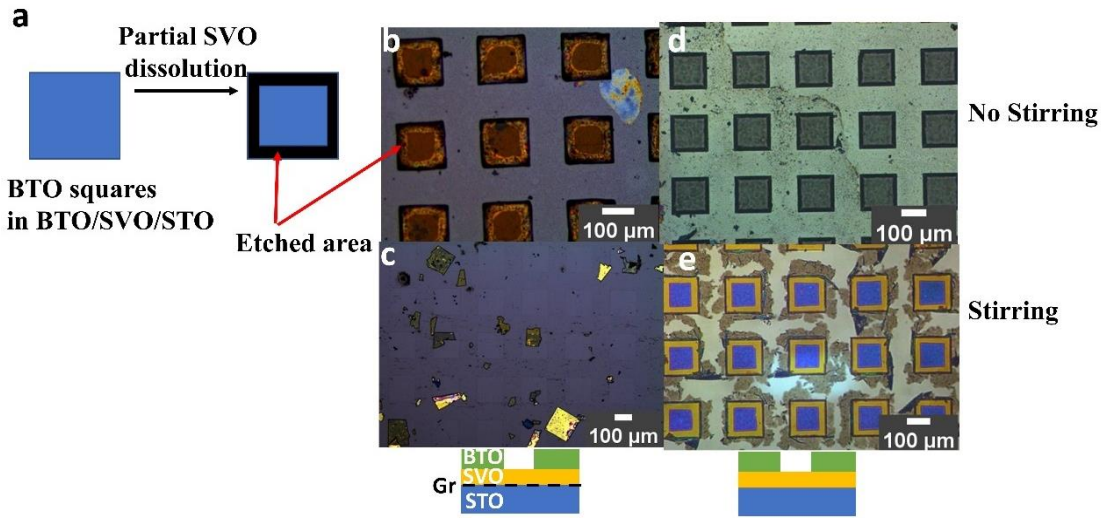

**Figure S9:** (a) Schematic of the etched area for dissolution rate determination, here,

$$Dissolution\ rates\ (\mu m^2/min) = \frac{Etched\ area\ (\mu m^2)}{Time\ dipped\ inside\ water\ (min)}$$

Comparison of the dissolution of SVO with Gr (b, c) and without Gr at the interface (d, e) when the solution is mildly stirred (c, e) vs. no stirring (b, d). Stirring the solution can improve the dissolution process. However, stirring causes the square to displace from its initial position for heterostructures with graphene-lined channels (sample A), as shown in (c). Therefore, stirring was avoided in this study.

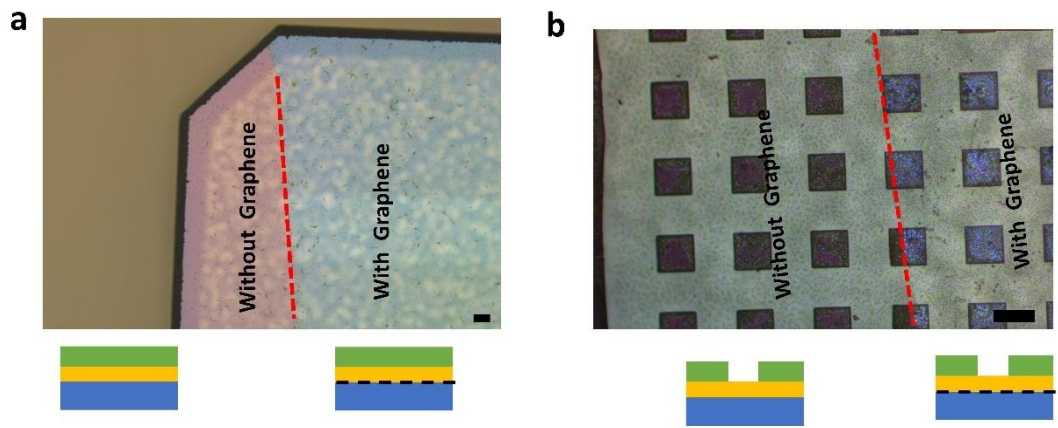

**Figure S10:** Correlating, an optical image of a BTO/SVO/Gr//STO heterostructure where Gr is partially transferred as shown in (a) with a graphene edge marked in red. After patterning the top BTO into squares and dissolving SVO in water optical image was captured (b). From the optical contrast, the BTO squares from the right (with graphene at the interface) released wholly compared to their counterpart squares on the left (without graphene at the interface), where there is minimal dissolution of SVO. Scale bar 100 μm.

**Comparison of membrane transfer through TRT vs PDMS, PPC/PDMS in stamping route:**

For PDMS, it was challenging to release the FS BTO on the target substrate; therefore, polymers such as poly(propylene) carbonate (PPC) coated PDMS sheets were used, facilitating the FS membrane's release process. The transparency of PPC/PDMS was better than TRT, enabling a deterministic transfer process. Mechanical force applied during the peeling step using TRT leads to more cracks as compared to transfer using PDMS.
Additionally, organic adhesive residue is observed while using TRT, as shown in Figure S11 (a, b). Also, the surface roughness is higher in the case of TRT than PDMS, as shown in Figure S11 (c, d). Hence, a clean and lower crack density transfer is achieved using PPC/PDMS handling layer compared to TRT while adopting the stamping route.

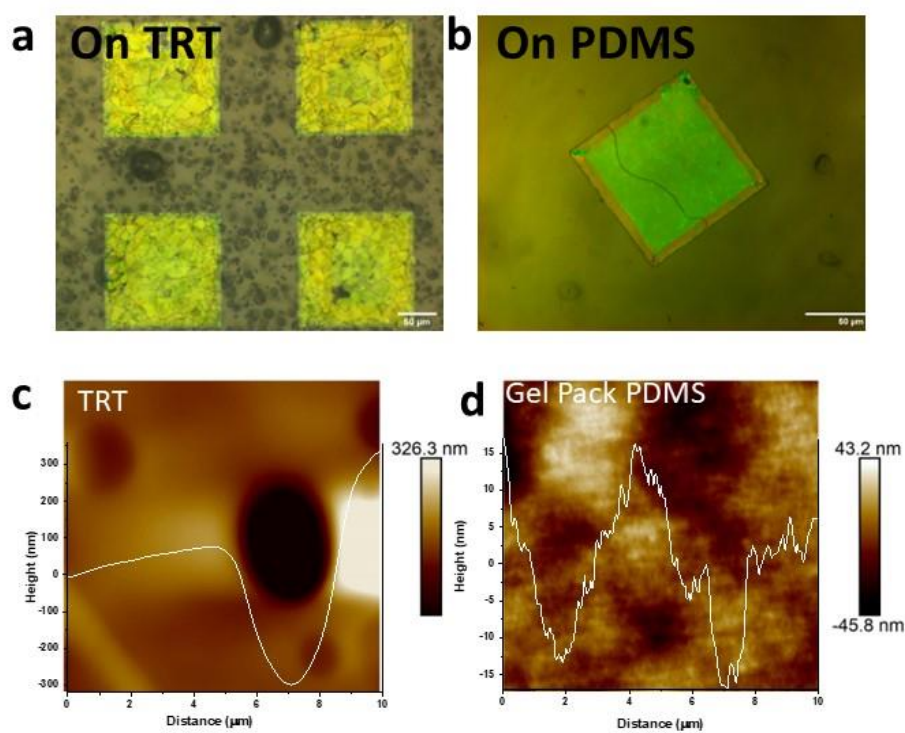

**Figure S11:** Square patterned BTO transferred onto (a) TRT and (b) PDMS. AFM micrograph of bare (c) TRT and (d) PDMS surface. Scale bar 50 μm.

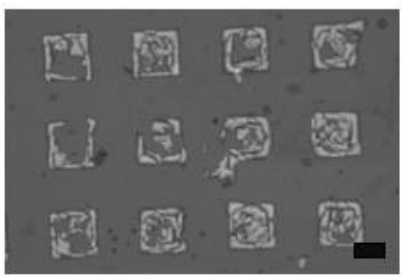

**Figure S12:** Transferred membrane through stamping route from reference sample without graphene at the interface.

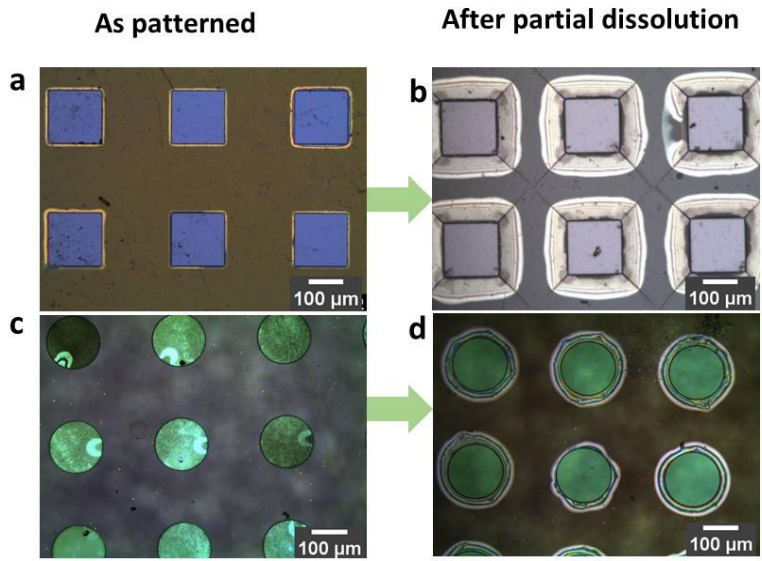

**Figure S13:** As patterned BTO with (a) square and (c) circular grids (left). (b-d) After the partial dissolution of SVO (right). Cracks propagate from the corners in square grids.

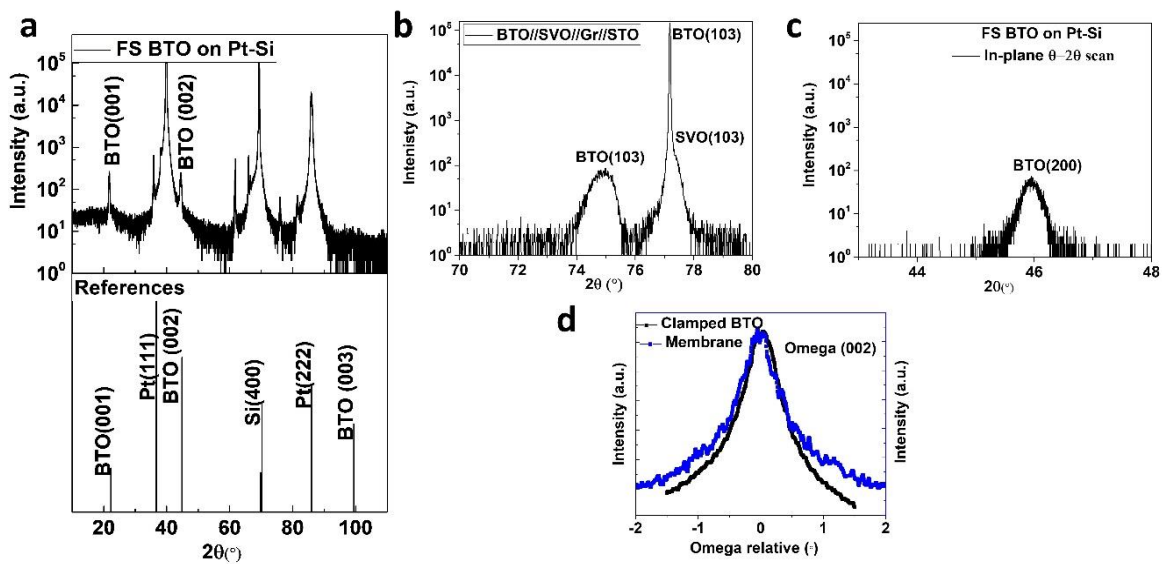

**Figure S14: XRD** (a) out of plane θ-2θ scan of FS BTO on Pt/Si from 10-120$^0$ with reference JCPDS data, (b) Asymmetric scan for (103) Bragg reflection of BTO in sample A (BTO/SVO/Gr//STO heterostructure), (c) In-plane scan for (200) Bragg reflection of FS BTO on Pt-Si substrate. (d) Omega scan for BTO (002) Bragg reflection for as-grown sample A (clamped) and after transfer onto Pt-Si substrate.

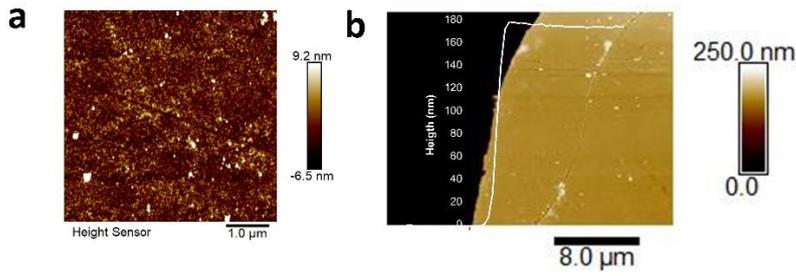

**Figure 15: AFM of transferred FS film.** (a) surface micrograph and (b) step height.

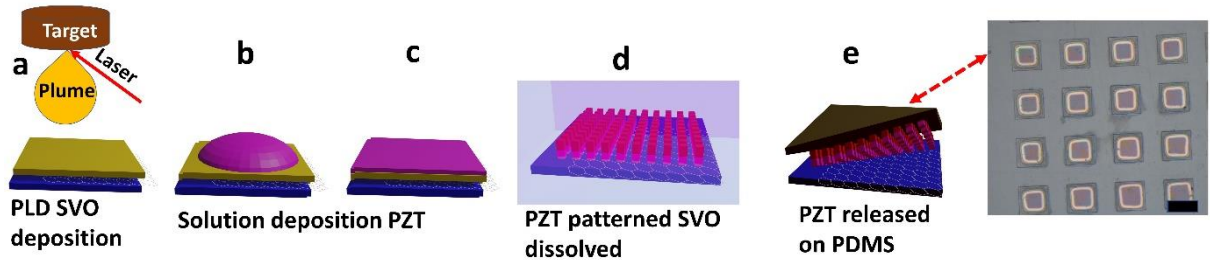

**Figure 16: Schematic representation of SVO/Gr//STO used as ex-situ template for solution deposition of PZT.** (a) PLD growth of SVO on Gr//STO, (b, c) chemical solution deposition of PZT, (d) square patterned PZT using lithography followed by wet etching (using $H_2O$: HCl: BHF of 4: 2: 1 and $HNO_3$: $H_2O$ of 3: 1 solution) and dissolution of underlayer SVO in water, (e) transfer onto PDMS and its corresponding optical image.

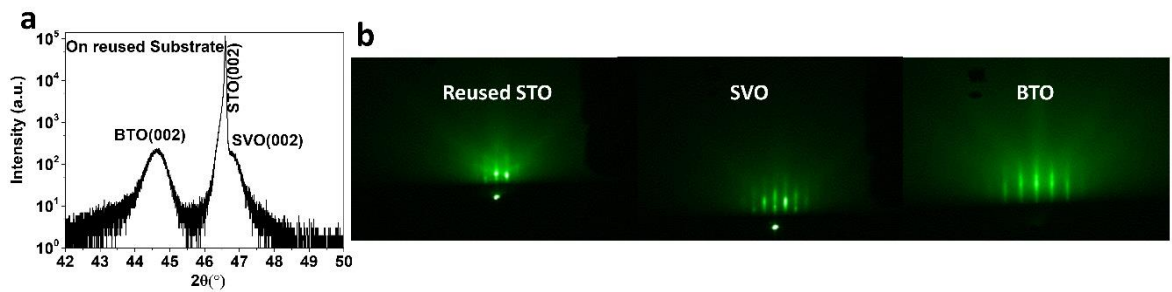

**Figure 17: Characterization of BTO/SVO on reused STO substrate:** (a) XRD out of plane θ-2θ scan around (002) Bragg reflection of BTO. (b) RHEED image of the reused STO substrate and as-grown SVO and BTO film.

**Video S1:** Video of the transfer process of BTO through the floating route. The target substrate used is Si.